\documentclass[12pt,preprint]{emulateapj}
\usepackage{apjfonts}

\begin{document}

\journalinfo{Accepted to the Astrophysical Journal Letters}
\submitted{Received 2004, Oct. 6; Accepted 2005 Jan. 21}

\title{A Discovery of Rapid Optical Flares from Low-Luminosity
Active Nuclei in Massive Galaxies}

\author{Tomonori Totani$^1$, Takahiro Sumi$^2$, George Kosugi$^3$, Naoki
Yasuda$^4$, Mamoru Doi$^5$, and Takeshi Oda$^1$} 

\altaffiltext{1}{
Department of Astronomy, Kyoto University, Sakyo-ku, Kyoto, 606-8502,
Japan} 

\altaffiltext{2}{Princeton University Observatory, Peyton Hall,
Princeton, NJ 08544-1001, USA } 

\altaffiltext{3}{Subaru Telescope,
National Astronomical Observatory of Japan, Hilo, HI 96720, USA}

\altaffiltext{4}{Institute for Cosmic Ray Research, The University of
Tokyo, Kashiwa, Chiba 277-8582, Japan}

\altaffiltext{5}{Institute of Astronomy, The University of
Tokyo, Mitaka, Tokyo 181-1500, Japan }



\begin{abstract}
We report a serendipitous discovery of six very low-luminosity active
galactic nuclei (AGNs) only by optical variability in one-month
baseline.  The detected flux variability is $\sim$ 1--5\% of the total
luminosity of host galaxies.  Careful subtraction of host galaxy
components in nuclear regions indicates that the fractional variability
$\Delta F/F$ of the nuclei is of order unity. At least one of them is
showing a compelling flaring activity within just a few days, which
appears to be quite different from previously known AGN variability. We
obtained spectroscopic data for the one showing the largest flare and
confirmed that it is in fact an AGN at $z = 0.33$ with an estimated
black hole mass of $\sim 10^8 M_\odot$.  As a possible interpretation,
we suggest that these activities are coming from the region around the
black hole event horizon, which is physically similar to the recently
discovered near-infrared flares of our Galactic nucleus. It is
indicated that our Galaxy is not special, and that surprisingly rapid
flaring activity in optical/near-infrared bands may be commonly hidden
in nuclei of apparently normal galaxies with low Eddington ratios, in
contrast to the variability of well-studied luminous AGNs or quasars.
\end{abstract}



\section{Introduction}  
Active galactic nuclei (AGNs) are generally variable, giving important
information on the activity of central super massive black holes
(SMBHs). In optical bands, most AGNs show significant variability on
time scales longer than months, but day-scale or shorter variability is
generally small (fractional amplitude of $\lesssim 10\%$) and rare (Ulrich,
Maraschi, \& Urry 1997; Webb \& Malkan 2000; Hawkins 2002). Therefore
the recent discovery (Genzel et al. 2003; Ghez et al. 2004) of
near-infrared (NIR) flares within one hour from Sgr A$^*$, an extremely
low-luminosity AGN of our Galaxy with a SMBH mass of $\sim 3 \times 10^6
M_\odot$, is surprising. Our Galaxy may be a peculiar AGN, or instead,
such variability may be common for very low-luminosity AGNs like Sgr
A$^*$, but it could not have been examined because of the difficulty in
observation for extragalactic sources.

In a deep search for faint transient objects using the Subaru
Telescope which has the largest field of view ($30'\times24'$) among 
existing 8m class telescopes, we serendipitously found several AGNs only
by variability on one month baseline, which is about 1--5\% of the total
luminosity of host galaxies. There have been several attempts to find
AGNs by their variability (Hawkins 1983; Bershady, Trevese, \& Kron
1998; Trevese et al. 1989; Sarajedini et al. 2003), but most of such
surveys looked for bright AGNs compared with host galaxy luminosity on
time baselines of many (five to ten) years. Therefore our discovery
is unique in terms of the time scale and nuclear/host luminosity ratios, 
providing us with a new method to find low-luminosity AGNs.
Here we report this discovery and also suggest a possible (though not
exclusive) interpretation.

\section{The Observation and Variable Object Search}

\begin{figure*}
\plotone{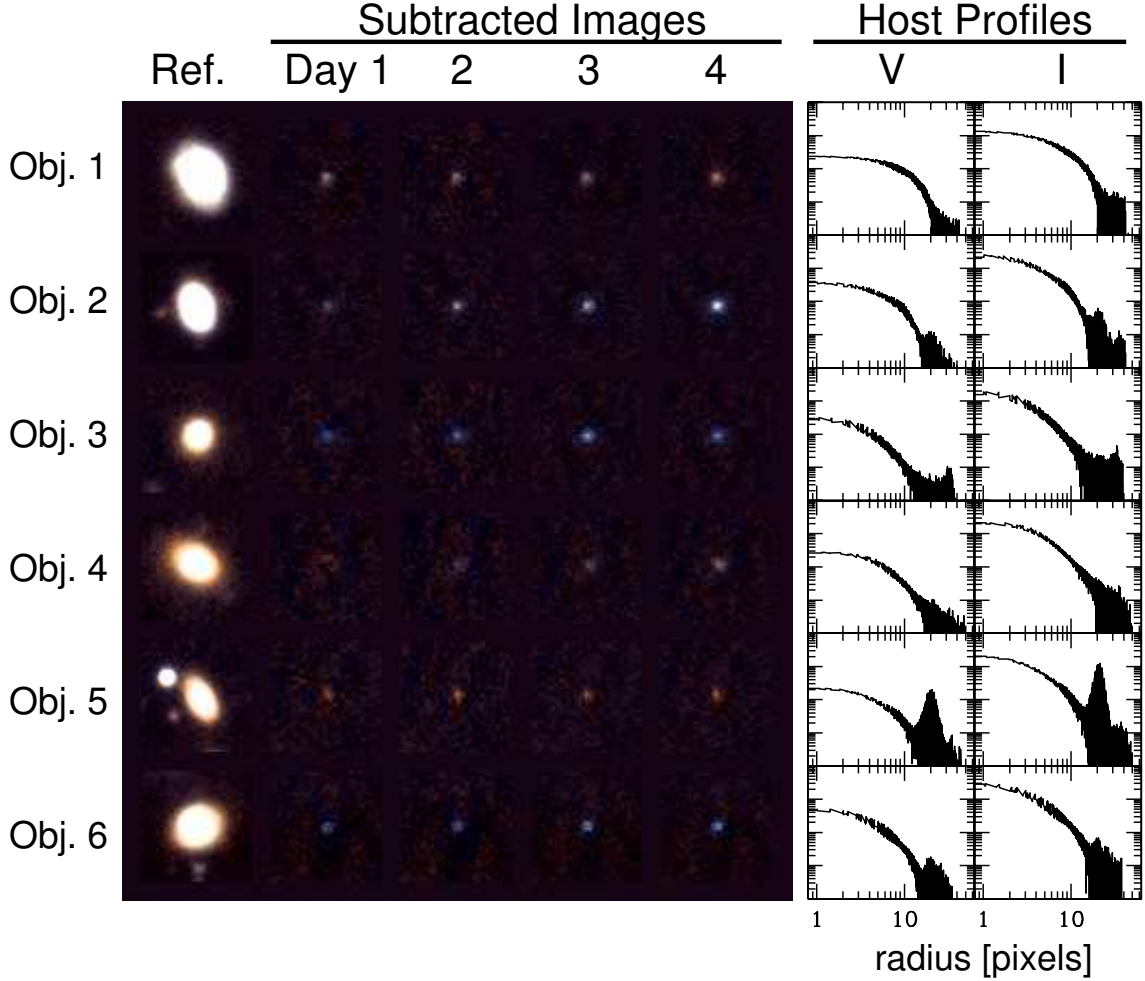} 
\caption{$V$ and $I$ color images of the six
variable AGNs found in the survey.  The first column shows the
reference images taken about one month prior to the main observing run,
and the reference-subtracted images of the consecutive four nights in
the main run are shown from the second to fifth columns.  The size of
each panel is $10'' \times 10''$, and the pixel scale is
$0''.2$/pixel. The upper four objects (1--4) brightened in a month,
while the lower two faded. The subtracted images of the lower two are
inverted for presentation. The $V$ and $I$ surface brightness profiles
of host galaxies are also shown in the right panels; photon counts are
in arbitrary unit and two-dimensional surface brightness data are
projected onto the primary axis of elliptical fitting.  }
\label{fig:VIimage}
\end{figure*}

A fixed field centered on the galaxy cluster Abell 2152 was consistently
monitored in $V$ and $I$ bands with similar exposure times in this
observation, with the primary scientific purpose of searching
microlensing events (Totani 2003).  A reference frame was observed using
a night on May 5, 2003, and the main-run observations during four
consecutive nights were performed about one month after the reference
observation (June 1--4).  The weather condition was good and seeing was
mostly stable; the effective seeing FWHM on the stacked images over each
night is 0.76, 0.68, 0.78, 0.56, and 0.62 arcsec for the reference frame
and the four nights of the main run, respectively.  After stacking these
frames for each night, we subtracted the reference frame from the
main-run frames by using the Alard \& Lupton (1998) algorithm.  We then
searched for any brightened objects in subtracted frames (and in their
inverted frames for faded objects) by a standard software, SExtractor
(Bertin \& Arnouts 1996), requiring $S/N>10$ of the SExtractor best
magnitude both in $V$ and $I$ bands (about 26.1 and 24.4 mag,
respectively) in either of the four nights. The CCD counts were saturated
in the
centers of the brightest stars and galaxies, and these regions were
removed from the analysis because the subtraction does not work
correctly.

Using this procedure we detected more than 20 objects, and most of them are
likely to be supernovae that are associated with faint host galaxies but
are offset from their centers. However, if we choose variable objects
associated with bright galaxies ($I < 20$), we found 6 variable
objects, all of which are located at the very centers of
well-resolved host galaxies (Fig.  \ref{fig:VIimage}), indicating that
they are very likely to be AGNs. Assuming that supernovae trace stellar
light, the chance probability of finding a supernova in the center of a
host galaxy within the angular resolution is about 3\%,
and hence the probability that all these six objects are supernovae by chance
is $\sim 2 \times 10^{-10}$.  The objects 1--4 brightened in one month,
but the other two faded.

\begin{deluxetable*}{cccccccccccccc}[h]
\tabletypesize{\scriptsize} 
\tablecaption{Properties of Variable AGNs and Thier Hosts} 
\tablewidth{0pt} 
\startdata 
\hline \hline &
\multicolumn{6}{c}{$V$ band} & & \multicolumn{6}{c}{$I$ band} \\
\cline{2-7} \cline {9-14} 
ID \ \ & $F_{\nu, \rm tot}$ & $F_{\nu,
c}$ & \ $F_{\nu, c-*}$ \ & $\Delta F_{\nu, c}$ & $\Delta F/F$ [\%] &
$P_d$ [\%] & \ \ & $F_{\nu, \rm tot}$ & $F_{\nu, c}$ & $F_{\nu, c-*}$ &
$\Delta F_{\nu, c}$ & $\Delta F/F$ [\%]& $P_d$ [\%] 

\\ \hline 

1 & 32.10 & 3.18 & $-$/0.15/0.21 & 0.34 (9.2$\sigma$) & $-$/227/162 & 84 &
& 87.98 & 12.79 & 0.70/1.79/2.20 & 1.50 (13$\sigma$) & 214/84/68 & 11
\\

2 & 22.87 & 4.59 & 0.44/0.81/0.96 & 1.41 (34$\sigma$) & 320/174/147 &
0.0 & & 72.15 & 20.47 & 3.58/5.47/6.38 & 1.71 (14$\sigma$) & 48/31/27 &
7.9 
\\

3 & 9.92 & 3.15 & 0.20/0.58/1.02 & 0.81 (21$\sigma$) 
& 405/140/79 & 67 & & 36.45
& 13.88 & 2.84/4.42/6.04 & 0.68 (6.2$\sigma$) & 24/15/11 & 4.2 
\\

4 & 15.83 & 3.47 & 0.05/0.45/0.53 & 0.44 (12$\sigma$)
& 880/98/83 & 0.0013 & &
71.52 & 17.95 & 0.90/3.26/4.15 & 1.26 (11$\sigma$) & 140/39/30 & 6.7 
\\

\hline 

5 & 9.20 & 2.39 & 0.25/0.54/0.60 & 0.13 (3.8$\sigma$)
& 52/24/22 & 46 & & 41.51
& 14.89 & 3.43/4.92/5.00 & 1.48 (13.8$\sigma$) & 43/30/30 & 29 
\\

6 & 19.83 & 4.96 & 0.27/0.89/1.56 & 0.36 (11$\sigma$)
& 133/40/23 & 23 & & 69.00
& 19.95 & 3.27/5.73/7.74 & 0.34 (3.6$\sigma$) & 10/5.9/4.4 & 46 
\\

\hline \enddata
\label{table:agn} 

\tablecomments{The host galaxy total flux ($F_{\nu, \rm tot}$), nuclear
fluxes within $1.2''$ diameter aperture in the reference frame
($F_{\nu, c}$), estimates of AGN
components after star-light subtraction ($F_{\nu, c-*}$), variability
flux ($\Delta F_{\nu, c}$, one month baseline, maximum among the four
nights), and the fractional variability ($\Delta F/F \equiv \Delta
F_{\nu, c} / F_{\nu, c-*}$) are shown. All fluxes are in $\mu$Jy. The
normal photometric error within the aperture is $\sim 0.01$ and 0.04
$\mu$Jy for $V$ and $I$ bands, respectively, while the errors used to
calculate $S/N$ of $\Delta F_{\nu, c}$ (shown in parentheses) 
include those associated with the image
subtraction procedure.  The three numbers separated by slashes for
$F_{\nu, c-*}$ and $\Delta F/F$ are those assuming three different
profiles of host galaxy light: the $r^{1/4}$ law, the exponential
profile, and constant surface brightness within the aperture,
respectively.  For $V$-band data of the object 1, star-subtracted
nuclear flux is negative when the $r^{1/4}$-law profile is used, and
hence it is not shown. Significance of day-scale variability is
indicated by $P_d$, which is 
the chance probability of getting the data from
a constant source during the four nights.  The objects
1--4 brightened from May to June, but the other two faded. 
}
\end{deluxetable*}

To estimate the statistical significance of these detections, we
measured an aperture flux ($1.2''$ = 6 pixels diameter) of centers of
all $\sim$ 1,000 galaxies with $I < 20$ on the subtracted images.  The
distribution is fit by a Gaussian, and the $S/N$ ratios of the maximum
variability flux within the same aperture among the four nights are
estimated by using this empirical noise level. The results are shown in
Table 1. All objects are statistically significant using a criterion of
$S/N > 10$ in at least one band.  Three of the six objects show this
month-like timescale variability at such high levels of significance in
both bands.  In principle, we cannot exclude a possibility that these
are false detections created by some systematic failure of image
subtraction. However, failures of subtraction, which often happen in
saturated regions, can easily be recognized by complicated and clearly
artificial image profiles. The saturation effect is sensitive to seeing,
and if these detections were created by saturation, the variability flux
of all objects should show a similar dependence on the observed date,
which is not actually observed. Therefore we consider that these
detections of one-month scale variability are reliable. As shown below,
the spectroscopy of the object 2 shows some AGN features, which provides
a further support for the significance of the variability detection.

\section{Flux Variability Estimates}

\begin{figure}
\epsscale{1.5}
\plotone{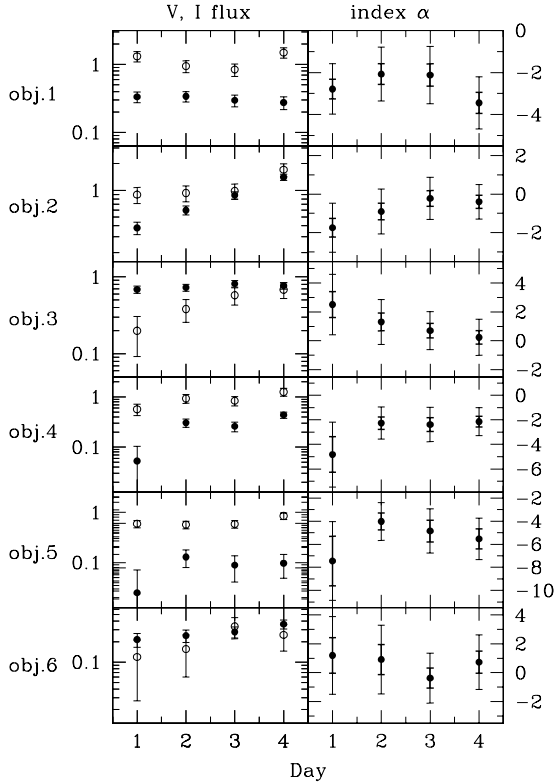}
\caption{(Left panels) Variability flux (in $\mu$Jy)
of the six AGNs during the four
nights of the main observing run, relative to the reference frame
taken one month earlier. 
The filled and open circles are for $V$ and $I$ bands,
respectively. 
(Right panels) Spectral index $\alpha$ ($f_\nu \propto
\nu^\alpha$) of variability flux between $V$ and $I$ bands.
The error bars in the left panel and thick error bars
in the right panel are statistical 1$\sigma$ errors, while the thin
error bars in the right panel are the total including systematics
(see text). 
The upper four objects brightened in a month, but
the lower two faded and their flux is measured in inverted subtracted
images.  
}
\label{fig:lc}  
\end{figure}

The fluxes of the variable objects measured on the subtracted images during
the four days are shown as light curves in Fig. \ref{fig:lc}.  The host
galaxy luminosities are shown in Table \ref{table:agn}, and the
variability flux is typically about 1--5\% of these.  We estimated
detection efficiency and photometric errors by placing artificial point
sources on the surface brightness peaks of the $\sim$1,000 galaxies with
$I < 20$. Then we repeated the same procedure of image subtraction and
source detection. We found that the detection efficiency is about 50\%
for $f_\nu = $0.176 and 0.515 $\mu$Jy, or 25.8 and 24.1 mag for $V$ and
$I$ bands, respectively.

We made two different error estimates: one is ``statistical'', which is
the variance from the mean of the observed flux of an artificial point
source during the four nights, and the other is ``total'' including
systematics, which is the variance from the real flux of an artificial
point source. The significance of variability within the four days
should be estimated by the statistical errors. The chance probability of
getting the data from a constant source during the four days is shown as
$P_d$ for each object in Table \ref{table:agn}. The evidence for
variability within the four days is compelling for the object 2 and
strong for the object 4. The spectral indices between $V$ and $I$ bands
are also shown in Fig. \ref{fig:lc}, with the total error estimates. The
evidence of spectral variability within days is weak for all
objects. The values of the indices show a large scatter from object to
object, but strong conclusions cannot be derived because of the large
systematic errors.

The AGN component in the pre-subtraction images must be estimated to
know the fractional variability ($\Delta F / F$) of the nuclei. There is
no apparent central excess in surface brightness profiles of host
galaxies (Fig. \ref{fig:VIimage}), indicating that stellar light
dominates AGN luminosity. We estimate the host galaxy component in the
nuclear regions of the reference frame as follows.  We set an aperture
diameter of $1''.2$ as the nuclear region. The surface brightness
profiles outside this aperture are elliptically fitted by the
$r^{1/4}$-law as well as the exponential profiles after convolved with
the seeing.  Then we subtract them from the observed total nuclear
fluxes, to get an estimate of the AGN components in the reference frame
(Table \ref{table:agn}).  The use of the exponential profile is
conservative because it gives lower host galaxy contribution and hence
lower fractional variability estimates, compared with the $r^{1/4}$-law
that is often used for the central stellar light profile. To be even
more conservative, we also estimate the host galaxy contribution by a
constant surface brightness within the aperture, with a value estimated
by photon counts along the aperture annulus. The dominant error is that
of variability flux $\Delta F$, which can be inferred from $S/N$ ratios
in the table.  These results indicate fractional variabilities ($\Delta
F / F$) of order unity for the six objects.

\section{Spectroscopy and Redshift Estimates/Determination}
Although there are two galaxy clusters at $z = 0.04$ and 0.13 in the
observed field (Totani 2003), the locations of the six AGNs are not
concentrated to the cluster centers.  The morphology of these host
galaxies is apparently elliptical or early-type, with magnitude of $I
\sim $ 18.5--19.5. The field galaxy counts at $I \sim 19$ dominate the
expected number of the cluster galaxies. According to a galaxy evolution
model that is in good agreement with various data (Totani \& Yoshii 2000),
the peak of the redshift distribution of early type galaxies at this
magnitude is $z \sim$ 0.3--0.4, and observed $V - I$ colors ($\sim$
1.6--2.2) are consistent with expected values for passively-evolving
early-type galaxies in this redshift range.  Therefore, it is likely
that these galaxies are field early-type galaxies at 
$z \sim$ 0.3--0.4, i.e.,  out of the two
clusters.

\begin{figure}[h]
\plotone{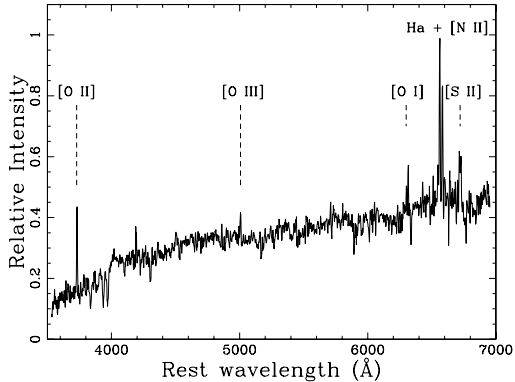}
\caption{The optical spectrum of the object 2. The redshift is $z=0.33$.}
\label{fig:obs2-spec}  
\end{figure}

We obtained an optical spectroscopic data by Subaru/FOCAS for the object
2 (Fig. \ref{fig:obs2-spec}), and confirmed that the redshift is, as
expected, $z = 0.33$ and the host galaxy is a typical giant galaxy with
$M_B \sim -20$.
Several emission lines are detected in the nuclear region, whose FWHMs
($\sim$ 300 km/s) are typical of narrow line regions.  The line ratios
are typical of the low ionization nuclear emission regions (LINERs),
confirming that this object is a low-luminosity AGN.  In the following,
we assume $z = 0.4$ for the other five AGNs for order-of-magnitude
discussions.  Using the empirical relation between bulge luminosity and
black hole mass (McLure \& Dunlop 2001), the SMBH mass of these AGNs
should be about $\sim 10^8 M_\odot$, and the nuclear luminosity becomes
$\sim 10^{-5}$--$10^{-4}$ in units of the Eddington ratio, i.e., $(\nu
L_\nu)/L_{\rm Edd}$, where $L_{\rm Edd}$ is the Eddington luminosity.

\section{Discussion}
For the first time, except for the recently discovered flares in Sgr
A$^*$, fractional AGN variability of $\sim$ 100\% within a few days in
optical/infrared bands is detected in one, and probably six, such
low-luminosity (low Eddington ratio) AGNs. The Sgr A$^*$ flares
have been interpreted as violent nonthermal phenomena 
at the innermost region of a radiatively inefficient accretion flow
(RIAF) around the SMBH (Yuan, Quataert, \& Narayan 2004). Such behavior
cannot be seen for luminous AGNs whose time variability has been studied
much better; the only exception is the rare population of blazars whose
jets are closely directed to the observer (Ulrich et al. 1997).
Although yet other interpretations might be possible, we here examine
these two interpretations (RIAFs and blazars).

The continuum-to-line luminosity ratio is a useful diagnostic to
discriminate between the two interpretations.  Since the continuum of
blazars is strongly beamed by relativistic jets, unbeamed emission lines
are generally weak and even difficult to see. On the other hand, if we
are observing less beamed continuum emission from accretion disks, we
expect a continuum-to-line ratio that is typical of normal (non-blazar)
AGNs. The H$\alpha$ emission ($2.6 \times 10^{40}$ erg/s) is clearly
detected for the object 2 and it argues against the blazar
interpretation. By using the empirical relations between line and
continuum luminosities (Ho \& Peng 2001; Wang, Staubert, \&
Ho 2002), the H$\alpha$ luminosity translates into the
expected optical continuum luminosity of $\nu f_\nu \sim 1.6 \times
10^{41} $ and $7.8 \times 10^{43}$ erg/s for normal AGNs (Seyfert 1s)
and blazars, respectively. This should be compared with the variability
flux of the object 2: $\Delta (\nu f_\nu) \sim$ (0.9--3.2) $\times
10^{42}$ erg/s in the $V$ band. The observed variability flux is
intermediate between the two interpretations, but considering that the
object 2 is likely in a flare phase, the mean flux appears to be closer
to the RIAF interpretation.

The number density of these six AGNs is about $\sim 10^{-4} \ \rm
Mpc^{-3}$, using the comoving volume out to $z = 0.4$. If the AGN
variability is from disk emission, this number can be compared with
published AGN luminosity functions in optical (Ho 2004) or
X-ray (Ueda et al. 2003) bands at a nuclear activity level of $ \nu L_\nu
\sim 10^{42}$ erg/s, which is found to be $\phi(>10^{42}{\rm erg/s})
\sim 10^{-3.5} \ \rm Mpc^{-3}$.  These numbers are not very different,
indicating that the six AGNs found in this work constitute a
considerable fraction of low-luminosity AGNs. On
the other hand, if variability is from blazar activity, the real
number density should be much larger than $\sim 10^{-4} \ \rm
Mpc^{-3}$, because of the collimation of blazar emission.  It is
believed that blazars are beamed by a factor of $\Delta \Omega /
(4\pi) \sim \Gamma^{-2}/4 \lesssim 1/400$, where $\Gamma \sim 10$ is
the jet Lorentz factor (Salvati, Spada, \& Pacini 1998), 
and hence the real number
density of AGNs like the objects found in this work 
should be $\gtrsim 10^{-2} \ \rm
Mpc^{-3}$.  However, the isotropic H$\alpha$ line luminosity indicates
that the nuclear activity must be larger than $\sim 10^{41}$ erg/s,
and the number density of such AGNs should not be larger than
$\sim 10^{-3} \ \rm Mpc^{-3}$ from the AGN luminosity functions. This
provides another argument against the blazar interpretation.

It is interesting to note that, if the variability reflects the scale of
emission region around SMBHs, the variability time scale of $\sim$ 1
hour for the $3 \times 10^6 M_\odot$ black hole of Sgr A$^*$ corresponds
to $\sim$ 1 day for $\sim 10^8 M_\odot$ black holes. Although the
Eddington ratio of our AGNs ($\sim 10^{-5}$--$10^{-4}$) is much larger
than that of the Sgr A$^*$ ($\sim 10^{-9}$), it is well below the border
($\sim 10^{-2}$) separating the standard thin disk and RIAFs (Kato,
Fukue, \& Mineshige 1998). A RIAF model of NIR/X-ray flares of Sgr A$^*$
predicts that flare activity should disappear with increasing Eddington
ratio, but it remains up to $L/L_{\rm Edd} \sim 10^{-5}$ in optical/NIR
bands (Yuan et al. 2004). Therefore it seems a reasonable interpretation
that previously unknown violent activity in optical bands has been
hidden around the SMBH event horizon of very low Eddington-ratio AGNs, which
cannot be seen in luminous AGNs whose luminosity is close to the 
Eddington limit.

We would like to thank S. Mineshige and the anonymous
referee for useful comments. 
This work has been supported by the Grant-in-Aid for the 21st
Century COE ``Center for Diversity and Universality in Physics'' from
the Ministry of Education, Culture, Sports, Science and Technology
(MEXT) of Japan.

\end{document}